\documentstyle[12pt,world_sci]{article}
\input BoxedEPS.tex
\SetRokickiEPSFSpecial
\input pslogos
\begin{document}

\title{{\bf VLPC CHARACTERIZATION FOR THE DZERO UPGRADE}}
\author{D. Adams, M. Adams, B. Baumbaugh, I. Bertram, A. Bross, D. Casey,\\
        S. Chang, M. Chung, C. Cooper, C. Cretsinger, R. Demina, G.
Fanourakis,\\
        T. Ferbel, S. Gr\"unendahl, J. Hinson, B. Howell, H. Johari, J. S.
Kang, \\
        C. L. Kim,  S. K. Kim, D. Koltick, F. Lobkowicz, S. Margulies, \\
        J. Moromisato, M. Narain, C. H. Park\thanks{Representing the D0
Collaboration.},
        S. Reucroft, Y. M. Park, R. Ruchti,\\
        J. Solomon, E. VonGoeler, J. Warchol, M. Wayne, E. Won, Y. Yu\\
 {\em Fermi National Accelerator Laboratory, University of Illinois at Chicago,
 }\\
 {\em Korea University, Kyungsung University, Northeastern University,}\\
 {\em University of Notre Dame, Purdue University, Rice University,}\\
 {\em University of Rochester, Seoul National University}\\}


\maketitle

\setlength{\baselineskip}{2.6ex}

\begin{center}
\parbox{13.0cm}
{\begin{center} ABSTRACT \end{center}
{\small \hspace*{0.3cm}
We have studied the performance
characteristics of approximately 4000 Visible Light Photon Counters(VLPC)
over a range of operating temperatures and biases. We will describe the
characterization procedure and present the measured gain, noise,
and relative quantum efficiency for the devices.
We will also discuss the general
operation of the devices and the significant problems encountered during the
characterization.}}
\end{center}

\section{Introduction}

\par In 1997, D\O ~will upgrade the central tracking region of the
detector with an entirely new system consisting of a 2 Tesla
superconducting solenoid, a 6000 channel preshower detector made of
scintillator
with wavelength shifting fiber readout, an outer 80000 channel
scintillating fiber tracker
and an inner 10$^6$ channel silicon strip tracker\cite{1}.
The optical signal
from the fibers will be detected by Visible Light Photon Counters(VLPCs).
VLPCs are arsenic doped silicon diodes that operate near a
temperature of 6.5 K with a bias of approximately 6.5 V\cite{2}.
The detailed examination of the characteristics of the VLPCs is the subject of
this report.

\section{Characterization Stand}

\subsection{Characterization Procedure}

\par The VLPC chips are produced by Rockwell
and are supplied as 1 $\times$ 8 pixel arrays with individual
pixels of 1 mm diameter.  The arrays are mounted on a ceramic substrate to form
a hybrid and this hybrid is then mounted on a Torlon carrier that allows for
fiber alignment with the pixels.  Noise, gain, and quantum efficiency (QE)
measurements were taken on all pixels at operating points between 5.5 and 7.0 K
and with VLPC bias between 5.9 and 7.1 V.  An LED light pulser was used to
inject an optical signal into the devices. Pulser data along with pedestal
data were used to calculate
noise, gain, and relative QE.

\subsection{Systematic Errors}

\par The combined systematic errors on the gain due to the run to run change
of cassette condition, the channel to channel variation of preamp gain
and the channel to channel VLPC bias variation is about 5\%.

\par The channel to channel light level variation is estimated by
cycling a hybrid (calibration chip) through each of the positions in
the cassette and measuring the response.  The systematic error on QE
measurement from light level variation is about 5\%.

\section{Results}

\par Figure 1 shows a typical spectra obtained from a good pixel.
One sees well separated pedestal, first and second photoelectron peaks.

\begin{figure}
\begin{center}
\vspace*{1cm}
\TrimTop{6 cm}
\TrimBottom{13cm}
\TrimRight{11cm}
\TrimLeft{5cm}
\HideDisplacementBoxes
\BoxedEPSF{goodspectra.ps scaled 800}
\caption{\tenrm A typical spectrum from a good VLPC pixel at T=6.5 K and V=6.5
V.}
\label{fig1}
\end{center}
\end{figure}

%
%

\subsection{Dark Current}

\par Dark current was measured by fitting the pedestal
spectra to a Gaussian distribution. The extra counts above
the tail were considered to be dark current from single
photoelectron events. The rates shown in Fig. 2 are essentially
the singles rates at a threshold of 0.5 photoelectrons.
%
%
\begin{figure}
\begin{center}
\vspace*{1cm}
\TrimTop{5 cm}
\TrimBottom{5 cm}
\HideDisplacementBoxes
\BoxedEPSF{dpf_plot_all.ps scaled 800}
\caption{\tenrm Dark current, Gain and QE as a function of VLPC bias
at T=6.5 K and as a function of operating temperature at V=6.5 V.}
\label{fig2}
\end{center}
\end{figure}

%
%

\subsection{Gain}

\par The gain is defined as the difference between the first and second
photoelectron peak positions. At our nominal operating point of 6.5 V and 6.5
K, the gain = 12.6 $\times$ 10$^4$ $\pm$ 2500.

\subsection{Quantum Efficiency}

\par The relative QE is obtained by comparing the average number of
photoelectrons observed from pixel to pixel and normalizing the
distribution to an average QE of 60\% at 6.5 K and 6.5 V.

\subsection{Yield}

\par 520 hybrids have been characterized. Of those characterized,
320 were considered good at the test stand by looking at spectra from
the T=6.5 K and V=6.5 V point, 120 had at least 1 bad pixel, and
80 were considered marginal and were retested. Of the marginal chips re-tested,
34 were recovered as good chips. The recovery can be attributed to better
contact with the read-out springs in the cassette.

\section{ Acknowledgements}

\par This research is being performed with the generous support of the
Department of Energy.

\bibliographystyle{unsrt}

\end{document}